\documentclass[aps,prx,showpacs,twocolumn,groupedaddress,amsmath,amssymb]{revtex4-1}
\usepackage{times,xspace}
\usepackage{amsfonts}
\usepackage{amsfonts}
\usepackage{bm}
\usepackage{amssymb}
\usepackage[final]{graphicx}
\usepackage{color}

\begin{document}

\title{Reconstructing the bulk Fermi surface and the superconducting gap properties
from Neutron Scattering experiments}

\author{Tanmoy Das$^{1,2}$, R. S. Markiewicz$^1$, and A. Bansil$^1$}
\affiliation{$^1$Physics Department, Northeastern University, Boston, MA, 87545, USA.\\$^2$Theoretical Division, Los Alamos National Laboratory, Los Alamos, NM, 87545, USA.}
\date{\today}

\begin{abstract} 
We develop an analytical tool to extract bulk electronic properties of unconventional superconductors through inelastic neutron scattering (INS) spectra. Since the spin excitation spectrum in the superconducting (SC) state originates from Bogoliubov quasiparticle scattering associated with Fermi surface nesting, its energy-momentum relation--the so called
`hour-glass' feature--can be inverted to reveal the Fermi momentum dispersion of the single-particle spectrum as well as the corresponding SC gap function. The inversion procedure is analogous to the quasiparticle interference (QPI) effect in scanning tunneling microscopy (STM). Whereas angle-resolved photoemission spectroscopy (ARPES) and STM provide surface sensitive information, our inversion procedure provides bulk electronic properties. The technique is essentially model independent and can be applied to a wide variety of materials.
\end{abstract}

\maketitle \narrowtext

The study of cuprate superconductivity has led to the discovery of a number of
unexpected relationships between seemingly different spectroscopies.
STM provides an outstanding example of this: while basically a real-space probe, it can
extract momentum-($k$)-space information of the electronic Fermi surface and SC pairing
symmetry usually associated with ARPES. Elastic scattering of Cooper pairs leads to the
QPI pattern measured by STM.\cite{hanaguri,kohsaka}  By analyzing the QPI pattern as a
function of scattering angle (in $q$-space) it is possible to reconstruct both band
structure and gap information in $k$-space.  While the technique was originally
developed for $d$-wave cuprates, it is finding wide applications in other materials,
including pnictides, where the SC gap is probably of $s_{\pm}$ symmetry, and in
topological insulators in the non-superconducting state.

The development of a similar technique for INS would have obvious advantages.  First,
ARPES and STM are surface sensitive probes, which in practice means that they can only
be performed on materials that are readily cleaved.  For the same reason, they are
mainly restricted to quasi-two-dimensional materials, and the question always remains
how sensitive the results are to surface effects, such as pinning, reconstruction, and
excess scattering.  In contrast, INS is a bulk probe which does not need special surface
preparation and can readily be applied to three-dimensional materials.  For instance,
INS results on heavy fermion materials were available years before the first
high-resolution ARPES studies.  Here we demonstrate that inelastic scattering between
the particle and hole Bogoliubov quasiparticles generates a similar magnetic
quasiparticle scattering (MQPS) profile which can be probed directly by INS
measurements.\cite{chubukov,norman,eremin}  Taking all three spectroscopies together, we are able to establish a
definite consistency between the $r-$space (STM), $k-$space (ARPES), and $q-$space
(INS) dynamics of the unconventional Bogoliubov quasiparticles.

The overall phenomenology of neutron scattering in cuprates is well-known, and a number
of universal features have been experimentally
identified\cite{wilson,vignolle,bourges,heyden,fong,woo,pailhes04,tranquada,daiybco}
and theoretically interpreted.\cite{chubukov,norman,eremin}
The results indicate that a distinct low energy magnetic mode is present near
the antiferromagnetic nesting vector ${\bm Q}=(\pi,\pi)$ in almost all the
cuprates.  The intensity of this mode is enhanced in the SC state while its
energy scales $\omega_{res} ({\bm Q}) \propto2\Delta_{SC}$ for all
cuprates\cite{yu,foot1}, suggesting a close connection of these modes with
SC pairing. The dispersion of spectral weight away from this resonance peak
also has a universal character, forming an `hour-glass'-like pattern centered on the
resonance mode and displaying a 45$^o$ rotation on passing through the
resonance peak. Below the resonance energy, spectral weight disperses along the
Cu-O bond direction, while above the resonance the dispersion peak lies along
the diagonal direction.  Despite an overall universality, many features of this
`hourglass' dispersion are highly material specific and we show that the difference comes mainly
from the nature of the pseudogap order among other band structure related properties.
\begin{figure}[here]
\centering
\rotatebox{0}{\scalebox{.53}{\includegraphics{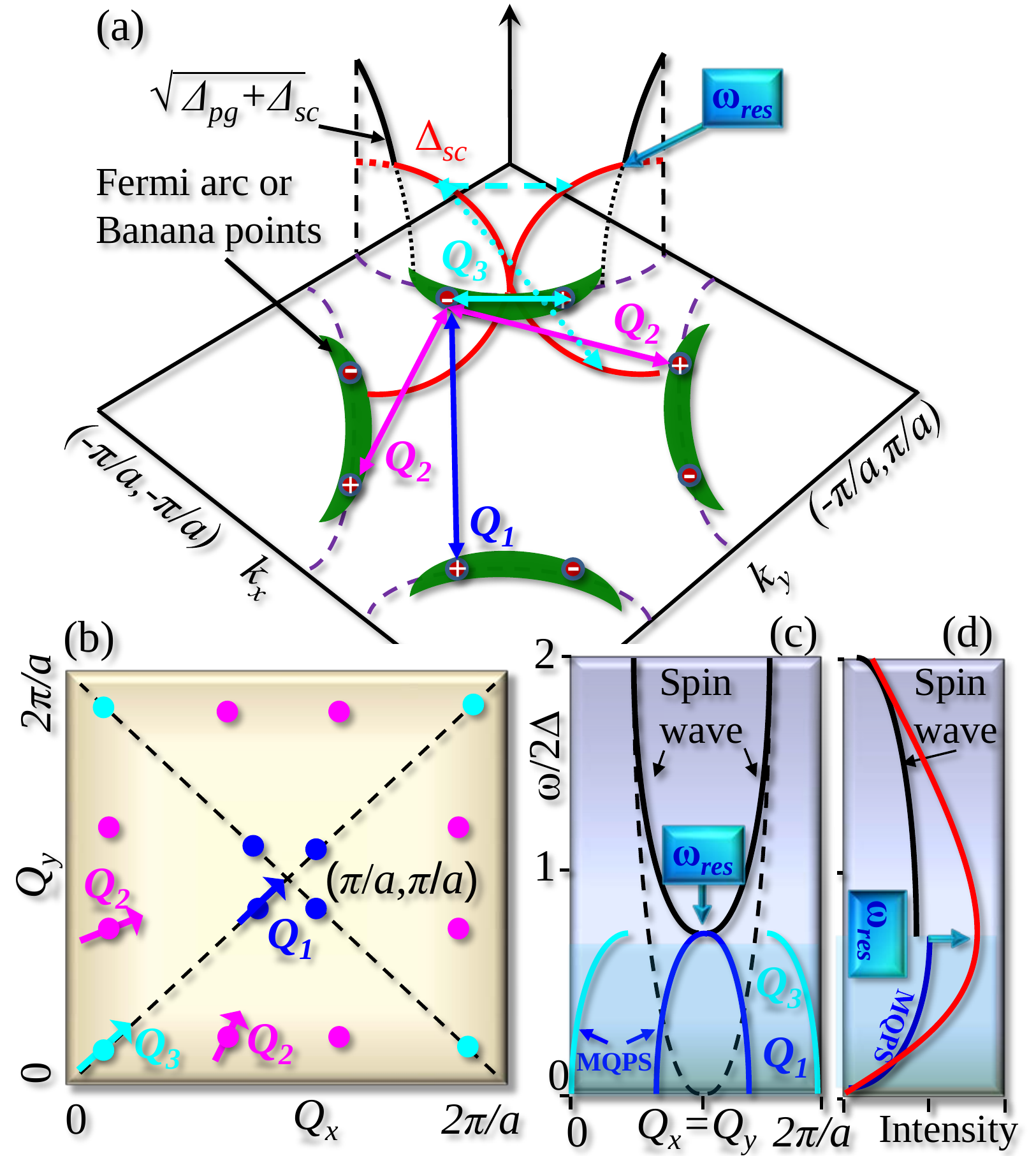}}}
\caption{{\bf Origin of MQPS vectors.} (a) Sketch of cuprate Fermi surface, showing four `banana'-shaped regions (green shapes) with intense spectral features at their tips (red dots).  The gap as a function of angle is shown on one banana.  Above the Fermi pocket, the superconducting gap is cut off by competing-order pseudogap (black solid line).
On this gap curve the contrasting behavior of QPI and MQPS vectors is illustrated: the QPI vectors connect two banana tips at the same energy, as shown by the long dashed cyan line,
while the MQPS vectors actually connect banana tips across the Fermi surface, from an electron tip at $\omega<0$ to a hole tip at $\omega>0$, shown by the dotted cyan arrow for the $Q_3$ branch.  For convenience the various MQPS vectors are defined by projected   horizontal arrows, labelled $Q_1-Q_3$, illustrating the two bright points that they connect.
(b) The $q$-space representation of the MQPS. Symbols
represent the $Q_i-$vectors of same color as
shown in (a). The associated arrows point in the direction that the $Q_i$
vectors shift with increasing energy along the Fermi surface. (c) The energy and momentum
dispersion of two $Q_i$ vectors are shown along the diagonal 
direction [$Q_x=Q_y$]. [The intensity of $Q_3$ is expected to be low (see text).]
The black solid line depicts the spin-waves which are gapped below the
SC gap while the black dashed line stands for the gapless spin-wave (schematic)
dispersion expected in the normal state. The black dashed line is the linear spin-wave
expected in the non-superconducting region. (d) Both spin-wave (solid black) and MQPS (solid blue)
have monotonic intensity variation as a function of energy, and as they meet at $Q$,
the total intensity (red line) attains a sharp peak.}
\label{fig1}
\end{figure}

We begin with describing the inversion procedure in INS,
which is analogous to the common inversion procedure performed using the QPI features in the STM spectra.
Fig.~1(a) depicts a typical form of the Fermi surface (green shadings)
which is assumed to be truncated below the antiferromagnetic Brillouin zone due
to the presence of underlying pseudogap ordering. When the superconducting gap has a $d-$wave form $\Delta_{\bm k}=\Delta_0(cos(k_xa)-cos(k_ya))/2$, where $a$ the in-plane lattice constant, the spectral function at any particular energy $\omega\le2\Delta_0$ has a maximal intensity at eight bright-points $K(\omega)$ that develop on four banana-pockets which satisfy $\omega=\Delta_{K(\omega)}$ (red dots). In STM experiments, elastic scattering at energy $\omega$ is dominated by scattering between these points, leading to a form of Friedel oscillations known as QPI, satisfying the condition\cite{hanaguri,kohsaka}
\begin{equation} \label{eq:0}
\omega=\Delta_{K(\omega)}=\Delta_{K(\omega)+q_i}.
\end{equation}
The long dashed cyan arrow in Fig.~1(a) illustrates one possible $q_i$-vector.

The INS experiments, which measure the imaginary part of the transverse spin-spin correlation function $\chi^{\prime\prime}$, can also be understood through an analogy with the QPI pattern.\cite{chubukov,eremin} Indeed, since $\chi$ is related to the joint density-of-states (JDOS), the neutron scattering will be dominated by transitions from banana tips below the Fermi level to banana tips above the Fermi level, as illustrated by the dotted cyan arrow in Fig.~1(a). The energy of a particular transition will be given by
\begin{equation} \label{eq:1}
\omega(Q_i)=|\Delta_{{\bm k}}|+|\Delta_{{\bm k}+{\bm Q}_i}|=\Delta_0\left[|g_{{\bm k}}|+|g_{{\bm k}+{\bm Q}_i}|\right],
\end{equation}
where the $d-$wave structure factor is $g_{\bm k}=(\cos{(k_x a)}-\cos{(k_ya)})/2$.

While the same banana points contribute to both QPI and neutron scattering, there are actually more ${\bm q}-$vectors in the latter case owing to more possibilities of inelastic scattering. In elastic scattering there are 7 $q_i-$vectors at any energy that connect a given banana point to the other banana points.\cite{hanaguri,kohsaka} However, in INS there is a coherence factor which allows only the poles for which $\Delta_{\bm k}$ and $\Delta_{{\bm k}+{\bm Q}_i}$ have opposite signs, since the magnetic neutron scattering cross-section is odd under time-reversal symmetry.\cite{chubukov,norman,eremin} Hence we identify four neutron scattering vectors which dominate in cuprates and give rise to a spin excitation profile, which can be called MQPS pattern, in analogy with the QPI patterns. The MQPS vectors can be denoted as ${\bm Q}_{1,2,3}\sim {\bm q}_{3,6,7}$, see Fig.~1(a), where the lower-case ${\bm q}$s are the corresponding QPI vectors.\cite{hanaguri,kohsaka}

There is a second, more important difference from QPI.  Note that for fixed $\omega$ Eq.~\ref{eq:0} is the equation of a point in $q$-space, while Eq.~\ref{eq:1} is the equation of a curve. Because there will be a different $Q_i$ for each pair $\Delta_{{\bm k}}$ and $\Delta_{{\bm k}+{\bm Q}_i}$ which satisfy Eq.~\ref{eq:1}.  However, we show below that
this is not a problem: the gap and FS can be reconstructed from INS data along any $Q-\omega$ cut. INS data are mostly available along the diagonal cut $Q_x=Q_y$, and along the bond direction, i.e. $[(0,\pi)\rightarrow(2\pi,0)]$ and equivalent cuts as a function of energy. We therefore define the spanning vectors as $Q_1$ and $Q_3$ along the diagonal direction and $Q^{\prime}_1$ and $Q^{\prime}_3$ along the bond direction [note that $Q_2$ does not lie on these two special cuts]. The diagonal cut is a particularly simple choice, since along this cut the MQPS vectors are exclusively associated with special points, for which $\Delta_k=-\Delta_{k+Q_i}$. For these special points the analogy with QPI becomes exact, with $\omega_{MQPS}= 2\omega_{QPI}$. Along the bond direction, the situation is similar as discussed below.

The $Q$-space positions of the poles of Eq.~2, or the MQPS pattern,  are plotted schematically at a constant energy in Fig.~1(b). The associated arrows indicate the direction each vector moves as the excitation energy increases which is simply a manifestation of the shape of the Fermi pocket and the $d-$wave superconducting gap values at each Fermi momentum. To see this, we draw an illustrative energy versus momentum dispersion
relation for two $Q_i$ vectors along the diagonal direction in Fig.~1(c).
Branch $Q_1$ disperses from $\omega=0$, where the incommensurate $Q_1$-vector connects pairs of nodal points, and hence measures the width of the hole pocket, to the resonance peak at $(\pi,\pi )$ at the maximum $\omega=2\omega_{res}$ connecting the hot-spots at the boundary
of the magnetic Brillouin zone -- that is, $Q_1$ is the dispersion branch associated with the
magnetic resonance peak.  Branch $Q_3$ represents an intrapocket scattering, and the corresponding branch starts from $Q_3=0$, $\omega=0$ at the nodal point, dispersing towards a $Q_3$-vector which spans the length of the hole pocket when $\omega=2\omega_{res}$.  Thus both curves attain the maximal $\omega$ when the banana
points correspond to the hot spots along the Brillouin zone diagonals -- the maximal diagonals of the hole pockets.
A similar phenomenon is found in the experimental QPI spectra.\cite{hanaguri,kohsaka} In summary, the inelastic MQPS intensity pattern at an energy $\omega =2\Delta$ in neutron scattering exactly corresponds to the elastic QPI pattern seen at an energy $\omega =\Delta$ in STM, and hence can also be used to reconstruct the Fermi surface and SC gap properties.

To illustrate how the above results play out in practice, we compare them to experiment and to realistic calculations of INS. We calculate the INS spectrum of the cuprates from a one-band Hubbard model, with self-energy corrections from a GW calculation called the quasiparticle-GW (QP-GW) model.\cite{tanmoyop,DasNFL} The QP-GW hamiltonian consists of four components which are calculated self-consistently: $H_{LDA}+H_{SDW}+H_{SC}+\Sigma$. The electronic dispersion, $H_{LDA}$, is based on a tight-binding fit to the material-specific first-principles single band dispersion of the antibonding combination of Cu $d_{x^2-y^2}$ and O $p_{x/y}$ orbitals.\cite{BobTB} The pseudogap state of the cuprates is treated as a spin-density wave (SDW), $H_{SDW}$, based on a Hubbard term in the Hamiltonian, which is treated using standard random-phase-approximation (RPA) theory.\cite{schrieffer} Below $T_c$, $d$-wave SC order develops $H_{SC}$ which couples naturally to the SDW state.\cite{Dasprl} The resulting energy spectrum is $E({\bm k})=\sqrt{(E^s)^2({\bm k}) + \Delta_{\bm k}^2}$, where the non-superconducting dispersion for both spin states are $E_{s}=\xi^+({\bm k})\pm E_{0}({\bm k})$ [for $E^2_0=\xi^-({\bm k})^2+G^2$, $\xi^{\pm}({\bm k})=(\xi({\bm k})+\xi({\bm k}+{\bm Q}))/2$, $\xi({\bm k})$ is the non-interacting band in the Bloch state, and $G$ is the effective SDW gap which produces the
pseudogap above the antiferromagnetic zone boundary]. The SDW gap causes a substantially reconstructed Fermi surface (FS) at low-temperature as shown in Fig.~2(c). At low temperature the $d-$wave superconducting gap coexists with the SDW state. By fixing $U$ and the pairing interaction $V$, to account for the experimental values of pseudogap and superconducting gap respectively (see Table I), there are no free parameters in the susceptibility calculation. Finally, we calculate the self energy  due to the magnetic and charge excitations which renormalizes the overall dispersion by a momentum independent mass renormalization.

The full susceptibility is computed in all spin channels within random phase approximation (RPA). In RPA, the resonance pole is determined by the real part of the Lindhard susceptibility, $\chi_0^{\prime}$ which attains a logarithmic divergence at all the Bogoliubov quasiparticle scattering vectors. Simultaneously $\chi_0^{\prime\prime}$ possess discontinuous jump due to Kramer's Kronig relation. In this spirit, the positive divergence in $\chi_0^{\prime}$ or the discontinuous jump in $\chi_0^{\prime\prime}$ can be used as an indicator of the observed the magnetic spectra in the SC state.
In the SDW state, the situation is more complicated, nevertheless, the overall phenomenology can be discussed approximately by the imaginary part of the transverse spin susceptibility which mainly consists of three important factors (see appendix for more details):
\begin{eqnarray}\label{eq:2}
\chi_0^{\prime\prime}({\bm
q},\omega)\approx\frac{\pi}{N}\sum\sum_{\bm
k}A({\bm k},{\bm q}) C({\bm k},{\bm q})
\delta(\omega-\omega_{res}({\bm k},{\bm q})).
\end{eqnarray}
Here, $\omega_{res}({\bm k},{\bm q})=E({\bm
k})+E({\bm k+q})$, which converges to Eq.~2 on the Fermi surface [$E^{s}({\bm k})=0$]. Since the INS spectrum is proportional to the RPA susceptibility $\chi^{\prime\prime}$, the associated intensity of a given pole in the spectrum is controlled mainly by the two coherence factors $A$ and $C$ associated with antiferromagnetic zone folding and SC gap symmetry breaking, which can be approximated for discussion as
\begin{eqnarray}
A({\bm k},{\bm q}) &=&\frac{1}{2}\left(1-\frac{\xi^{-}({\bm k})\xi^{-}({{\bm k}+{\bm q}})+G^2}{E_{0}({\bm k})E_{0}({\bm k}+{\bm q})}\right),\\
C({\bm k},{\bm q}) &=&\frac{1}{2}\left(1-\frac{E^{s}({\bm k})E^{s}({\bm k}+{\bm q})+\Delta_{\bm k}\Delta_{{\bm k}+{\bm q}}}{E(\bm k)E({\bm k}+{\bm q})}\right).
\end{eqnarray}
At the normal state Fermi surface [$E^{s}({\bm k})=0$], the superconducting coherence factor reduces to $C=1/2(1-{\rm sgn}(g_{\bm k}){\rm sgn}(g_{{\bm k}+{\bm q}}))$ which attains its maximum value of 1 whenever $\Delta_{\bm k}$ and $\Delta_{{\bm k}+{\bm q}}$ have opposite signs, thereby explaining why the spectrum is dominated by the four magnetic scattering channels $Q_i$ with $(i=1-4)$. Furthermore, at small $|q-Q|$, the SDW coherence factor, Eq.~5, simplifies\cite{sachdev} to $A\rightarrow 1-(\xi^{-}({\bm k})+\xi^{-}({\bm k}+{\bm q}))G^2/4E^4_0({\bm k})=1-{\mathcal{O}}((Q-q)^2)$, which attains its maximum value of 1 at $q=Q$. This explains why the intensity for $Q_1$ branch being closer to $Q$ dominates over the other branches and the intensity gradually increases to its maximum value at $Q$.

\begin{figure}[here]
\rotatebox{0}{\scalebox{.55}{\includegraphics{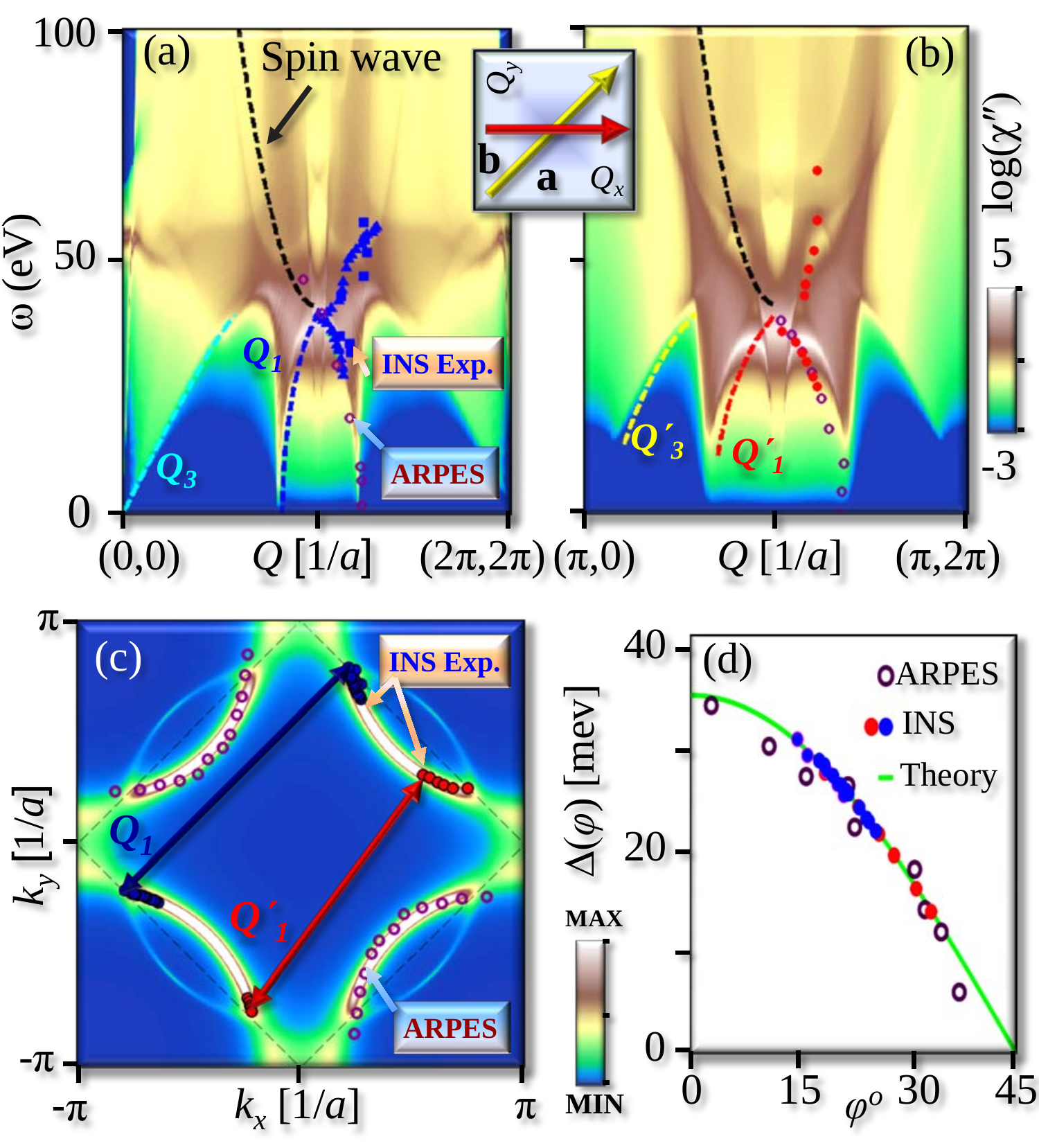}}}
\caption{{\bf Reconstruction of Fermi surface from the INS spectra.}
(a)-(b) Computed $\chi^{\prime\prime}(q,\omega)$ is plotted along the
diagonal and bond directions respectively (see inset) in the superconducting
state of YBCO. The experimental data for YBCO$_{6+y}$ are shown in various symbols [green stars ($y=0.85$) =Ref.~\cite{bourges}, blue squares ($y=0.85$) =Ref.~\cite{pailhes04}, blue triangles
($y=0.95$)=Ref.~\cite{reznik}]. The data along the bond direction in (b) are obtained for $y=0.6$ (red symbols) from Ref.~\cite{hinkov}. The blue and cyan dashed lines in (a) are the plot of Eq.~2 for $Q_1$ and $Q_3$ (red and yellow dashed lines are for corresponding equivalent vectors $Q_1^{\prime}$ and $Q_3^{\prime}$ along bond directions in (b). The brown symbols give the resonance spectrum calculated from Eq.~2 using ARPES dispersions\cite{yagi} for $y=0.6$ as input [see (c)]. The experimental and theoretical lines are only plotted in one side of two equivalent directions for clarity. We plot $\chi^{\prime\prime}(q,\omega)$ in log-scale to highlight weaker features. The black line is a guide to the eye for the dispersion expected for gapped spin-waves, which constitutes the upward dispersion in the resonance spectra.
(c) Plot of computed Fermi surface of YBCO. The blue symbols are the locus of the Fermi momenta determined from the experimental neutron data shown in (a) solving Eq.~2 while red symbols are the same but along the bond direction shown in (b). The brown symbols are the ARPES Fermi surface from Ref.~\cite{yagi}. All symbols have four-fold symmetry, but are plotted here only along one particular quadrant for visualization. The blue and red arrows are the scattering vectors $Q_1$ and $Q_1^{\prime}$.
(d) The extracted superconducting gap plotted as a function of Fermi surface angle [0$^o$ being the antinodal direction and 45$^o$ the nodal direction] from the neutron data of (a) and (b), compared to the ARPES data (brown symbols) of Ref.~\cite{nakayama} and the present theoretical curve.} \label{md2d}
\end{figure}

Fig.~2 presents our main result, showing that the dominant spectral features in neutron
scattering correspond exactly to the MPQS features, in a SDW superconductor.
For a specific example we analyze YBa$_2$Cu$_3$O$_{6+y}$ (YBCO).
The color plots in Fig.~2(a) and 2(b) show two-dimensional $q-\omega$ intensity maps of the calculated imaginary part of the RPA spin susceptibility ($\chi^{\prime\prime}$) along the diagonal and the bond directions for YBCO at $y=0.85$. Shown superimposed on the left sides of Fig.~2(a) are the calculated MQPS dispersions, with $Q_1=(\pi\pm\delta,\pi\pm\delta)$ plotted as a dashed blue line and $Q_3=(\pm\delta,\pm\delta)$ as a dashed cyan line, computed from Eq.~2. Similarly, in Fig.~2(b) the dashed red [yellow] line represents the equivalent branch along the bond direction: $Q^{\prime}_1=(\pi,\pi\pm\delta)/(\pi\pm\delta,\pi)$ [$Q^{\prime}_3=(\pi,\pm\delta)/(\pm\delta,\pi)$]. The susceptibility maps present complex patterns with numerous dispersing features. Nevertheless the MQPS branches $Q_1$ and $Q_3$ capture the essential shape of the two downward dispersions, while a spin wave branch is clearly seen dispersing upward away from the magnetic resonance feature. 

There are two main sources of intensity variation for each $Q-$vector. As discussed earlier, the SDW coherence factor is strongly momentum dependent and reaches maximum at ${\bm Q}$. Since branch $Q_3$ stops dispersing well before it reaches ${\bm Q}$, its intensity is relatively low while $Q_1$ branch gains more intensity as it moves towards the resonance. The equivalent $Q_1^{\prime} (Q_3^{\prime})$ along the diagonal direction in Fig.~2(b) has twice as large intensity as $Q_1 (Q_3)$ in Fig.~2(a). This is due to an overall degeneracy factor. The scattering vectors $Q^{\prime}_{1,3}$ have one commensurate direction and one incommensurate one while $Q_{1,3}$ are incommensurate along both $x-$ and $y-$directions except at the resonance at ${\bm Q}$. Thus $Q'_{1,3}$ connect twice as many Fermi surface points as any other $Q_{1,3}$. As a result the intensity of the magnetic spectrum along the bond direction [in Fig.~2(a)] is twice as large as that along the diagonal direction [in Fig.~2(a)]  [also see several intensity plots at relevant momentum cuts in supporing Fig.~S4]. Therefore, in the superconducting region, the intensity profile is rotated along the bond-direction as shown in Fig.~3(a). We find good agreement with the experimental results, represented by symbols of various colors\cite{bourges,pailhes04,reznik,hinkov}, on the right-hand side Figs~2(a) and 2(b).

However, here our principal task is to demonstrate that the Fermi surface and superconducting gap can be reconstructed from the experimental data. The lower branch of the experimental data which disperses downward from $(Q,\omega_{res})$ is associated with the $Q_1$ scattering vectors. Therefore, by solving Eq.~2 at the experimental points, we determine a map of $k_F$ which agrees very well with the theoretical Fermi surface as well as with the ARPES FS.  Note that we find the same Fermi surface by utilizing the diagonal cut $Q_1$ [blue dots in Fig.~2(c)] or the bond direction $Q_1^{\prime}$ [red dots]. For a cross-check, we use the experimental FS measured by ARPES\cite{yagi} [brown symbols in Fig.~2(c)] to compute the magnetic resonance spectra by solving Eq.~2, brown symbols in Fig.~2(a). This agrees well with the INS data. Furthermore, the extracted values of $k_F$ versus $\omega$ can be used to determine the underlying superconducting gap symmetry using Eq.~2 which assumes $d_{x^2-y^2}-$symmetry.  This also agrees well with ARPES data\cite{nakayama} and the present theory, see Fig.~2(d). The superconducting gap can be extracted up to the edge of the magnetic Brillouin zone or the hot-spot [$\phi\sim15^o$], above which the pseudogap dominates in the spectra and the MPQS features can no longer be followed.

Above we have discussed the magnetic resonance peak and the lower half of the hourglass.  The upper, rotated half of the hourglass has a different origin, related to spin waves. The black dashed line in Fig.~1(c) shows the spin-wave dispersion in the non-superconducting state at half-filling, which disperses away from the antiferromagnetic wave vector ${\bf
Q}=(\pi,\pi )$\cite{schrieffer}.  When superconductivity is turned on, low energy states near the Fermi surface are gapped (black solid line) and the paramagnon features at energies $\omega<2\Delta_0$ are washed out.  The black dashed lines in Figs.~2(a),~2(b) show that the present calculations capture this gapped spin wave spectrum.

\begin{figure*}[top]
\rotatebox{0}{\scalebox{.7}{\includegraphics{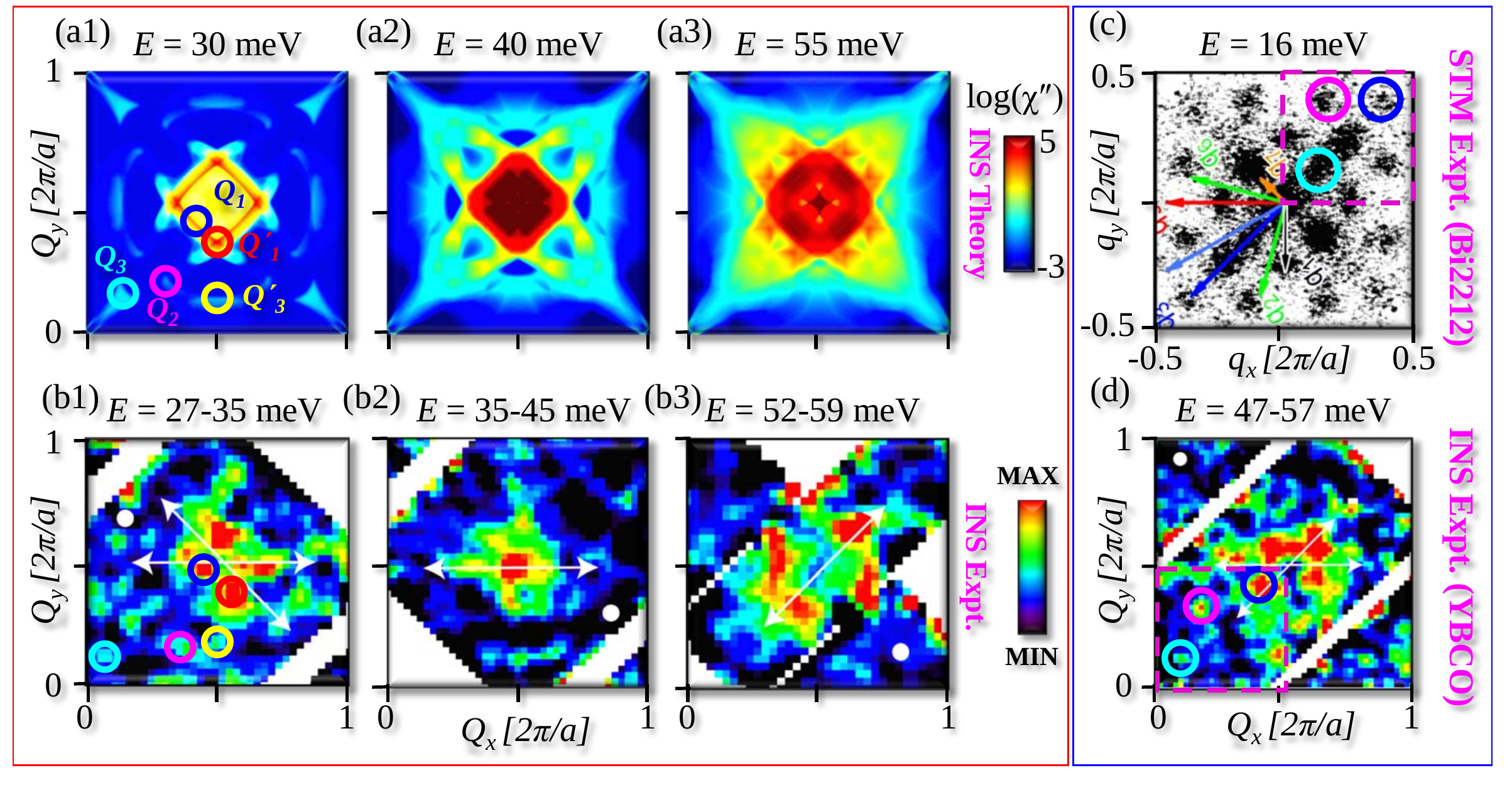}}}
\caption{{\bf Comparison of MQPS maps with experimental QPI maps. }
(a1)-(a3) Computed spectra of $\chi^{\prime\prime}(\omega)$ are plotted in color
code for three representative constant energy cuts, below the resonance,
near the resonance and above the resonance. (b1)-(b3) The corresponding experimental data
of YBCO$_{6.95}$ from Ref.~\cite{woo} for the acoustic channel. The white arrows in the experimental curves were used in the discussion in Ref.~\cite{woo}. (c) A QPI map of the STM data\cite{kohsaka} for Bi2212 is compared with the MQPS map of INS in (d). The MQPS spectrum is chosen in (d) for the optical channel to ease comparison with QPI maps as the magnetic profile is rotated here along the diagonal direction.
We note that QPI are usually plotted in the range [$-\pi$ to $\pi$] while the neutron scattering is plotted in the range [$0$ to $2\pi$]; the dashed box in (c), (d) represents the range [$0$ to $\pi$] to be compared. In (a1), (b1) and (d), circles of various colors depict various MQPS vectors defined before while the circles with same colors in (c) compare QPI vectors with corresponding MQPS vectors.} \label{md2d}
\end{figure*}

Combining the MQPS and spin wave spectra, Fig.~3 compares the $\omega$-evolution of our present theoretical MQPS maps [Fig.~3(b)] with corresponding experimental data [Fig.~3(c)]. At low energies the spectra are dominated by the MQPS vectors, and the figure also compares a QPI map [only available for Bi2212], in Fig.~3(c), with a neutron scattering map of the odd channel in YBCO, in Fig.~3(d).  Three high-symmetry vectors $Q_{1,2,3}$ (and two equivalent vectors along bond direction $Q_{1,2}^{\prime}$) are visible which correspond to the $q_{3,6,7}$ vectors in the QPI map [denoted by circles of the same colors]. In the low-energy region, the magnetic scattering profile of Fig.~3(a1) is squarish with finite intensity along all directions. However, the intensity is larger along the bond-directions than along the diagonal as discussed above and thus the profile is rotated along the bond directions, consistent with experiments shown in the corresponding lower panel. With increasing energy, at the resonance [Fig.~3(a2)] the intensity piles up at $(\pi,\pi )$ with tails dispersing towards $q=(0,0)$ which were recently seen in Hg-based compounds.\cite{yuan} Above the resonance, Fig.~3(a3), the magnetic spectrum is purely spin-wave based and the profile is rotated along the diagonal direction, as found experimentally.

In the above calculations, we have assumed that the superconductivity couples to the SDW order. When, the spin-wave spectrum of upward dispersion and the MQPS of downward dispersion meet at the commensurate antiferromagnetic vector $Q=(\pi/2,\pi/a)$, a resonance peak in the intensity occurs. Such a resonance peak is absent in a paramagnetic ground state.\cite{chubukov,norman,eremin} The SDW coherence factors play an important role in distributing the intensity over the entire magnetic spectrum, Eq.~4. Furthermore, the QPI-MQPS correspondence is obscured when the non-superconducting state becomes paramagnetic. While the $Q_i$-vectors still play a significant role, the magnetic resonance peak is shifted to a lower energy at a pole of the dynamic susceptibility\cite{eremin}.  In the magnetic superconducting ground state, the susceptibility peaks correspond to the MPQS $Q_i(\omega)$, with $\omega$ given by Eq.~2, as shown in Fig.~2{\it A},2{\it B}.

In summary, we have shown that while the spin-wave spectrum of collective mode origin persists at all dopings both in electron and holed doped cuprates including at half-filling, it becomes gapped in the low-energy region $\omega<2\Delta$ where the spectrum is dominated by the Bogoliubov scatterings of Cooper pairs.  When, the spin-wave spectrum of upward dispersion and the magnetic scattering of downward dispersion meet at the commensurate antiferromagnetic vector $Q$, a resonance peak in the intensity occurs. The INS spectroscopy is the only probe which can be utilized to reconstruct the Fermi surface and superconducting properties of the actual bulk ground state. Our method of analysis is independent of any particular model and can be performed entirely from experimental inputs. Therefore, this inversion procedure will also be useful in further research to extract the bulk-Fermi surface topology and pairing symmetry information in newly discovered iron-selenide,\cite{Dasfese,Dasvacancy} pnictide, chalcogenide\cite{Daspnictide} and heavy fermion\cite{DasHF} superconductors in which this information is still not settled. We also predict that the present formalism can be used to detect the electron pocket on the cuprate Fermi surface, if they are present in the bulk ground state.

\begin{acknowledgments}
This work is supported by the U.S.D.O.E grant DE-FG02-07ER46352, and benefited
from the allocation of supercomputer time at NERSC and
Northeastern University's Advanced Scientific Computation Center
(ASCC).
\end{acknowledgments}

\begin{appendix}

\section{Susceptibility in SDW+$d$-wave SC state}

\begin{table*}[top]
\centering
\begin{tabular}{|c|c|c|c|c|c|c|c|c|c|c|}
\hline \hline
Material& Doping ($x$) &$\Delta_{pg}$ (meV)& $U/t$ & $\Delta_{sc}$ (meV)& Pairing Potential & $T_c$ K & $\omega_{res}$\\
& & (Exp./Theory) & (Theory)& (Exp./Theory) & $V$ (meV) (Theory)& Exp.(Theory) & Exp. (Theory)\\
 \hline
 NCCO &  0.15& 60 [Ref.~\onlinecite{greven_NCCO}] & 4.1 &
 3.5 [Ref.~\onlinecite{qazilbash}] & -26 & 24 (37) [Ref.~\onlinecite{qazilbash}]& 4.5 (5) [Ref.~\onlinecite{greven_NCCO}]\\
 YBCO$_{6.85}$ & 0.21 &50 [Ref.~~\onlinecite{huefner}] & 2.5 &
 35 [Ref.~\onlinecite{reznik}] & -98 & 92 (160)  [Ref.~\onlinecite{reznik}] & 41 (40) [Ref.~\onlinecite{bourges}] \\
\hline \hline
\end{tabular}
\caption{Order parameters of the model. Experimental gap values and the resulting self-consistent values of the order parameters. Given the order parameters, theoretical values of $\omega_{res}$ are calculated and compared with experiment. The value of $U/t$ (where $t$ is nearest-neighbor hopping parameter) is chosen to reproduce the experimental peudogap ($\Delta_{pg}$) whose values are presented here along the `hot-spot' direction in the electron doped case and the antinodal direction for all hole doped cuprates. Similarly, the parameter value of pairing potential $V$ is taken to reproduce the superconducting gap ($\Delta$), whose maximum value along the antinodal direction is presented here.\cite{Das_nonmonotonic} Our mean-field calculations overestimate the values of $T_c$, presumably due to the neglect of phase fluctuations.\cite{Bob_sc}}
\label{Tab:ch2_order}
\end{table*}

Since the SDW state causes a unit cell doubling, the correlation functions
(Lindhard susceptibilities) are tensors in momentum
space representation\cite{schrieffer} We define the susceptibilities as the standard linear
response functions
\begin{eqnarray}\label{Eq:ch3_chi_1}
%
\chi^{ij}({\bm q},{\bm q}^{\prime},\tau)&=&
\frac{1}{2N}\Big\langle T_{\tau}\Pi_{{\bm q}}^i(\tau)\Pi^j_{-{\bm q}^{\prime}}(0)\Big\rangle
%
\end{eqnarray}
where the response operators ($\Pi$) for the charge and spin density
correlations respectively are
\begin{eqnarray}\label{Eq:ch3_rho_s}
\rho_{\bm q}(\tau)&=&\sum_{{\bm k},\sigma}
c_{{\bm k}+{\bm q},\sigma}^{\dag}(\tau)c_{{\bm k},\sigma}(\tau),\hspace{0.1in}{\rm and}~\nonumber\\
%
 S^i_{{\bm q}}(\tau)&=&\sum_{{\bm k},\sigma,\gamma}c_{{\bm k}+{\bm q},\sigma}^{\dag}
(\tau)\sigma^i_{\sigma,\gamma}c_{{\bm k},\gamma}(\tau).
%
\end{eqnarray}
The $\sigma^i$ represent two dimensional Pauli matrices along $i^{th}$
direction. $c_{{\bm k},\sigma} (c^{\dag}_{{\bm k},\sigma})$  is the destruction (creation) operator of an electronic state at momentum ${\bm k}$ and spin $\sigma$. For transverse spin response $S^{\pm}=S_x\pm iS_y$ whereas
longitudinal fluctuations are along the $z-$direction only.  In the present
($\pi,\pi$)-commensurate state, charge- and longitudinal spin-fluctuations
become coupled at finite doping. In common practice the transverse, longitudinal spin and charge
susceptibilities are denoted as $\chi^{+-}, \chi^{zz}$ and $\chi^{00}$
respectively. We collect all the terms into a single notation as
$\chi^{\sigma\bar{\sigma}}$ where $\bar{\sigma}=\sigma$ gives the charge and
longitudinal components and $\bar{\sigma}=-\sigma$ stands for the transverse
component. For the pure SDW state Eq.~\ref{Eq:ch3_chi_1} can be evaluated
rigorously. Here we generalize earlier calculations\cite{schrieffer} for realistic cuprate band structures. For the
combined SDW+$d-$SC state
%
%
\begin{eqnarray}
\chi_{ij}^{\sigma\bar{\sigma}}({\bm q},\omega)
&=&\frac{1}{N\beta}\sum_{{\bm k},n,s}
G_{is}({\bm k},\sigma,i\omega_n)G_{sj}({\bm k}+{\bm q},\bar{\sigma},i\omega_n+\omega)\nonumber\\
\label{chi1}\\
&=&\frac{1}{N}\sum_{{\bm k},\nu\nu^{\prime}}^{\prime}A^{\sigma\bar{\sigma}}_{\nu\nu^{\prime},ij}
\sum_{m=1}^3 C^{m}_{\nu\nu^{\prime}}\chi^{m}_{\nu\nu^{\prime}}({\bm k},{\bm q},\omega).
\label{chi2}
\end{eqnarray}
%
We obtain Eq.~\ref{chi2} from Eq.~\ref{chi1} after performing the Matsubara summation over $n$.
$G$ is the $4\times 4$ single-particle Green's function  in the Nambu space, constructed from the Hamiltonian given in the main text. The summation indices $\nu (\nu^{\prime}) = \pm$ gives two split SDW bands. Here, the coherence factor due to SDW order in the particle-hole channel is,
\begin{eqnarray}\label{chiSDW}
A^{\sigma\bar{\sigma}}_{\nu\nu^{\prime},11/22}&=&\frac{1}{2}\left(1\pm\nu\nu^{\prime}\frac{\xi_{\bm k}^-\xi_{{\bm k}+{\bm q}}^-+\sigma\bar{\sigma}(US)^2}{E_{0{\bm k}}E_{0{\bm k}+{\bm q}}}\right),\nonumber\\
%
%
A^{\sigma\bar{\sigma}}_{\nu\nu^{\prime},12/21}&=&-\nu\frac{G}{2}\left(\frac{\sigma}{E_{0{\bm k}}}+\nu\nu^{\prime}\frac{\bar{\sigma}}{E_{0{\bm k}+{\bm q}}}\right).
%
%
\end{eqnarray}
The SC coherence factors are
\begin{eqnarray}\label{t11}
&C^{\rm 1}_{\nu\nu^{\prime}}
=\frac{1}{2}\left(1+\frac{E^{s,\nu}_{\bf{k}}
E^{s,\nu^{\prime}}_{\bf{k}+\bf{q}}+\Delta_{\bf{k}}\Delta_{\bf{k}+\bf{q}}}
{E^{\nu}_{\bf{k}}E^{\nu^{\prime}}_{\bf{k}+\bf{q}}}\right),\nonumber\\
&C^{\rm 2/3}_{\nu\nu^{\prime}}=\frac{1}{4}\left(1\pm\frac{E^{s,\nu}_{\bf{k}}}{E^{\nu}_{\bf{k}}}
\mp\frac{E^{s,\nu^{\prime}}_{\bf{k}+\bf{q}}}{E^{\nu^{\prime}}_{\bf{k}+\bf{q}}}
%
-\frac{E^{s,\nu}_{\bf{k}}E^{s,\nu^{\prime}}_{\bf{k}+\bf{q}}+\Delta_{\bf{k}}\Delta_{\bf{k}+\bf{q}}}
{E^{\nu}_{\bf{k}}E^{\nu^{\prime}}_{\bf{k}+\bf{q}}}\right).
\label{sccoherence}
\end{eqnarray}
Lastly the index $m$ represents the summation over three polarization bubbles
related to the quasiparticle scattering ($m=1)$, quasiparticle pair
creation ($m=2$) and pair annihilation ($m=3$), as defined by
\begin{eqnarray}
\chi^{1}_{\nu,\nu^{\prime}}({\bm k},{\bm q},\omega)&=&
-\frac{f(E^{\nu}_{{\bm k}})-f(E^{\nu^{\prime}}_{{\bm k}+{\bm q}})}
{\omega+i\delta+(E^{\nu}_{{\bm k}}-E^{\nu^{\prime}}_{{\bm k}+{\bm q}})},
\label{Eq:ch3_chi_0_sc1}\\
\chi^{2,3}_{\nu,\nu^{\prime}}({\bm k},{\bm q},\omega)&=&\mp
\frac{1-f(E^{\nu}_{{\bm k}})-f(E^{\nu^{\prime}}_{{\bm k}+{\bm q}})}
{\omega+i\delta\mp(E^{\nu}_{{\bm k}}+E^{\nu^{\prime}}_{{\bm k}+{\bm q}})}.
\label{Eq:ch3_chi_0_sc23}
\end{eqnarray}
%

In the RPA model, the $2\times2$ susceptibility is obtained from the standard formula\cite{schrieffer}
\begin{widetext}
\begin{eqnarray}\label{RPASpinSucscomp}
\chi_{RPA,11}^{\sigma\bar{\sigma}}({\bm q},\omega)
 &=&\frac{\bigl[1+\sigma\bar{\sigma} U\chi_{22}^{\sigma\bar{\sigma}}({\bm q},\omega)\bigr]\chi_{11}^{\sigma\bar{\sigma}}({\bm q},\omega) +
  U\bigl[\chi_{12}^{\sigma\bar{\sigma}}({\bm q},\omega)\bigr]^2}
{\bigl[1-U\chi_{11}^{\sigma\bar{\sigma}}({\bm q},\omega)\bigr]\bigl[1+\sigma\bar{\sigma}U\chi_{22}^{\sigma\bar{\sigma}}({\bm q},\omega)\bigr]
+\sigma\bar{\sigma}\bigl[U\chi_{12}^{\sigma\bar{\sigma}}({\bm q},\omega)\bigr]^2},
\label{RPASpinSucscomp11}\\
\chi_{RPA,22}^{\sigma\bar{\sigma}}({\bm q},\omega)
 &=&\frac{\bigl[1-U\chi_{11}^{\sigma\bar{\sigma}}({\bm q},\omega)\bigr]\chi_{22}^{\sigma\bar{\sigma}}({\bm q},\omega) + U\bigl[\chi_{12}^{\sigma\bar{\sigma}}({\bm q},\omega)\bigr]^2}
{\bigl[1-U\chi_{11}^{\sigma\bar{\sigma}}({\bm q},\omega)\bigr]\bigl[1+\sigma\bar{\sigma}U\chi_{22}^{\sigma\bar{\sigma}}({\bm q},\omega)]
+\sigma\bar{\sigma}\bigl[U\chi_{12}^{\sigma\bar{\sigma}}({\bm q},\omega)\bigr]^2},
%
\label{RPASpinSucscomp22}\\
\chi_{RPA,12/21}^{\sigma\bar{\sigma}}({\bm q},\omega)
 &=&\frac{\chi_{12}^{\sigma\bar{\sigma}}({\bm q},\omega)}
{\bigl[1-U\chi_{11}^{\sigma\bar{\sigma}}({\bm q},\omega)\bigr]\bigl[1+\sigma\bar{\sigma}U\chi_{22}^{\sigma\bar{\sigma}}({\bm q},\omega)]
+\sigma\bar{\sigma}\bigl[U\chi_{12}^{\sigma\bar{\sigma}}({\bm q},\omega)\bigr]^2}.
%
\label{RPASpinSucscomp12}
\end{eqnarray}
\end{widetext}
In the longitudinal and charge channel ($\bar{\sigma}=\sigma$), the RPA corrections do not introduce any pole and thus all the normal structure lies above the charge gap in the particle-hole continuum. Along the transverse direction ($\bar{\sigma}=-\sigma$), a linear spin-wave dispersion develops in the normal state which extends to zero energy at $Q=(\pi,\pi)$.\cite{schrieffer} The necessary condition to yield a gapless Goldstone mode is that Eqs.~\ref{RPASpinSucscomp11}-\ref{RPASpinSucscomp12} reduce to the self-consistent SDW order parameter, $G$ at $q=Q$, which is indeed the case in the normal state. 

In the SC state, this zero energy spin-wave shifts to $\omega=|\Delta_{{\bm k}_F}|+|\Delta_{{\bm k}_F+{\bm q}}|$, due to the particle-particle (and hole-hole) scattering terms $\chi^{2,3}$ in Eq.~\ref{Eq:ch3_chi_0_sc23}. These terms have finite intensity only if the SC gap changes sign at the `hot-spot' ${\bm q}$,\cite{chubukov} see Eq.~\ref{sccoherence}. Above the SC gap, the spin-wave term coming from Eq.~\ref{Eq:ch3_chi_0_sc1} is turned on. The crossover between them creats the `hor-glass' pattern presented in Fig.~2.

\section{Mechanism of MQPS}
\begin{figure}[top]
\rotatebox{0}{\scalebox{.4}{\includegraphics{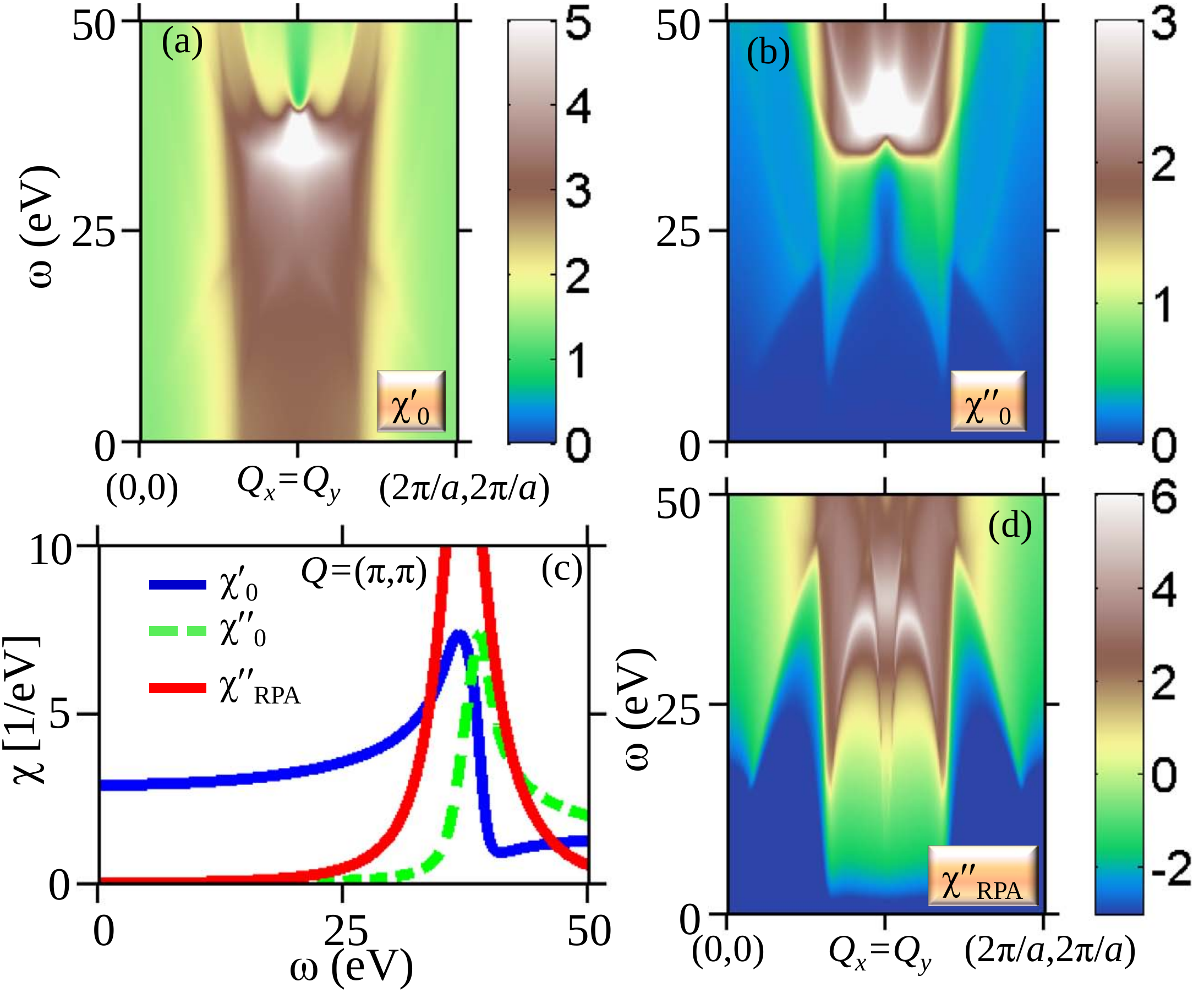}}}
\caption{The real and imaginary part of bare susceptibility χ0 is shown in (A) and (B) respectively for YBCO. (D), Corresponding RPA result plotted in a logarithmic intensity scale as in the case of Fig. 2 of the main text. (E), All three susceptibilities are shown at the commensurate momenta cut} \label{md2d}
\end{figure}

The result shown in Fig.~2 of the main text is obtained from a coexisting state of SDW and $d-$wave SC order within RPA framework. To see how one can use the behavior of imaginary part of the bare susceptibility as an indicator for the observed resonance behavior, we decompose the resonance spectra of Fig.~2(b) into its bare components as shown in Fig.~4. As shown in Fig.~4(a) and 4(b), the real and imaginary part of the bare susceptibility are related to each other by Kramers’ Kronig relation. Where the real part obtains a logarithmic divergence, see blue line in Fig.~4(c), the corresponding imaginary part possesses a discontinuous jump at the same location [green dashed line in Fig.~4(c)]. Within RPA, a resonance is possible when $\chi'_0=1/U$ condition is satisfied. In the region where $\chi'_0$ is greater than zero and also attains a divergence, a resonance can occur for a large range of $U$. In this spirit, a true resonance spectrum within RPA can simply be identified by tracing the divergences in $\chi'_0$ or by tracking the sudden peaks in $\chi''_0$. We emphasize that this argument holds even for multiband pnictide supercondcutors.\cite{Daspnictide}     

\end{appendix}

\end{document}